\documentstyle[12pt,aasms4]{article}

\received{ }
\accepted{  }
\journalid{ }{ }

  \slugcomment{   }

\begin{document}

\title{The Age of Gl879 and  Fomalhaut}

\author{David Barrado y Navascu\'es\altaffilmark{1,2}} 
\affil{Real Colegio Complutense, 26 Trowbridge St.,  Cambridge,
MA 02138, USA}

\author{John R. Stauffer\altaffilmark{3} and Lee Hartmann}
\affil{Harvard--Smithsonian Center for Astrophysics,
       60 Garden St., Cambridge, MA 02138, USA}

\and

\author{Suchitra C. Balachandran\altaffilmark{3}}
\affil{Deparment of Astronomy, University of Maryland, College Park, MD 20742}

\altaffiltext{1}{Present address: Center for Astrophysics,
    60 Garden Street, Cambridge, MA 02138}
\altaffiltext{2}{Depto. de Astrof\'{\i}sica. 
Universidad Complutense de Madrid, Av. Complutense s/n, 28040
Madrid, Spain.} 
\altaffiltext{3}{Visiting Astronomer, Cerro Tololo Inter-American
Observatory. 
CTIO is operated by AURA, Inc.\ under contract to the National
Science
Foundation.} 

1997, ApJ 475, 313

\begin{abstract}

     We estimate here the age of one of the prototypes
of the Beta Pic-like stars, Fomalhaut, based on the properties of
its common proper motion companion Gl879.  By combining
constraints
derived from the lithium abundance, rotational velocity, HR
diagram
position, and coronal activity we conclude that the age for
Gl879,
and hence the age for Fomalhaut, is 200$\pm$100 Myr.   
This age estimate agrees
quite well with the completely independent 
 age estimate derived directly from isochrone--fitting
to Fomalhaut's position in an HR diagram, and thus confirms that
circumstellar
dust disks can persist in A stars for several hundred Myr.

\end{abstract}


\section{Introduction}

One of the most interesting discoveries from the IRAS satellite
was the
detection of circumstellar dust disks around some nearby main
sequence
stars.  The prototypes for this class of stars were Beta Pic, Vega
and
Fomalhaut (Gillett 1986), and these disks are often referred to
as Beta Pic
or Vega disks.  As a result of their proximity and brightness,
these
three stars have been studied in most detail (Backman \& Paresce
1993), including
estimates
of the mass, radial distribution and structure, dust-grain
properties,
etc.  Subsequent to the discovery of the three
prototype objects, other groups have searched the IRAS database
and
identified a large number of less prominent members of the class
(Aumann 1985; Backman and Gillett 1987; Walker and Wolstencroft
1988).    
It has been estimated that 18\% of the field A stars
have
circumstellar dust disks with $\tau$ $\geq$\  2 x 10$^{-5}$ 
(Backman and Paresce 1993).  

It is generally believed that these circumstellar dust disks are 
either the direct descendents of T Tauri disks or the secondary
products of the planet formation process.
Knowledge of the ages of Beta Pic stars is therefore
one of the keys to the understanding of the formation and
evolution of
their disks.   However, to date there is little firm information
on the
ages of these systems.  No member of the class has yet been
detected in 
an open cluster.   Zuckerman et al. (1995) have estimated quite
young
ages ($<$ 10 Myr) for several Beta Pic stars, but in
most cases
these estimates are based on relatively qualitative indicators. 
All 
three of the prototypes are A stars, and estimates from their
post-ZAMS
evolution give rough ages of 100, 200, and 400  Myr 
for Beta Pic, Fomalhaut and Vega, respectively, with 
uncertainties about 30\% (Backman and Paresce 1993).
This age estimate for Beta Pic differs considerably from the
$\sim$2  Myr
proposed by Jura et al. (1993) based on the frequency of the A
stars with
$\tau$ $\geq$\ 10$^{-3}$ among the Bright Star Catalog  (Hoffleit
and Jaschek, 1993), and with the $\sim$10 Myr age estimated
for it by Lanz et al (1995) based on a re-evaluation of its position
in the HR diagram using an estimate of its surface gravity provided by UV
spectra.
However, we note that the Lanz et al. age has very large error
bars since the derived surface gravity differs at only the one
sigma level from the ZAMS gravity, and thus an age of 100 Myr
is also compatible with their data at the one sigma level.

As noted by Jura et al. (1993), the presence of a low mass
companion to one
prominent Beta Pic star - HR4796A - offers the chance to
establish a relatively
secure estimate for that star by determining a PMS
isochrone-fitting age
for the secondary.   By doing this, Jura et al. derived an age of
3 Myr for HR4796A.
Stauffer et al. (1995) obtained a high resolution spectrum to
determine the lithium
abundance of the secondary and also calibrated the 
isochrone--fitting age
by use of photometry for low mass stars in open clusters and
derived a revised age of 8$\pm$2 Myr for HR4796.

In this paper, we will report a new age estimate for one of the
three
prototypes - Fomalhaut - based on analysis of the properties of
its common-proper
motion companion Gl879.

\section{Observations and Data Reduction}

Table 1 lists some properties of Fomalhaut (Gl881) and Gl879.
 The photometric 
data were selected from Bessel (1990), the activity indicators 
were obtained from Panagi \& Mathioutakis (1993) and the IR
fluxes from
Oudmaijer et al. (1992) and Mathioudakis \& Doyle (1993).

As can be seen, the agreement between either the radial
velocities
and  apparent motion on the sky 
(Poveda et al. 1994) or the galactic velocity components
(Anosova \& Orlov 1991) indicates that both stars are, quite
probably,  physically
 associated, as was  originally suggested by Gliese (1969).
 Since the possibility of a capture is extremely low, 
we conclude that they must have been physically associated at
birth and therefore, it
is legitimate  to estimate the age of Fomalhaut
using constraints derived from its low mass companion.

A spectrum of Gl879 was obtained on May 21, 1995 with
the echelle
spectrograph on the Cerro Tololo Interamerican Observatory  4m
telescope.
We used the red, long camera  and a Tektronix 2048$\times$2048
CCD, a 31.6 l/mm grating and a 0.8 arcsec wide slit, yielding
a resolving power of about R$\approx$50000 at 
LiI 6708 \AA, as measured  with a Th--Ar comparison lamp. 
The total spectral range was $\lambda\lambda$5650--8050 \AA.
Standard bias subtraction, flat--field correction and
wavelength calibration was carried out  using  IRAF\footnote{
IRAF is distributed 
by National Optical Astronomy Observatories, which is operated by
the Association of Universities 
for Research in Astronomy, Inc., under contract to the National
Science Foundation, USA   }.  
The final  signal--to--noise ratio of the spectrum is $\sim$130.

Figure 1a shows the order which contains the LiI 6707.8 \AA{ } doublet.  This
feature is separable from the CN and FeI blend at 6707.4 \AA{ } 
due to the high  resolution of
the spectrograph and the low rotational velocity of Gl879 
(vsini$\le$4 km~s$^{-1}$,
estimated from the rotational period and the average radius for its spectral type).
The equivalent widths measured by fitting two Gaussian curves to
the spectrum are: EW(LiI)=33$\pm$2 m\AA{ } and EW(FeI+CN)=15$\pm$2 m\AA.
We also measured EW(H$\alpha$)=754$\pm$16 m\AA{ } in absorption.

There is in the literature a previous measurement
 of the LiI equivalent width in
Gl879.  Favata et al. (1995) obtained a value of EW(LiI)=35 m\AA.  Using
Pallavicini et al.'s (1987) curves of growth and T$_{\rm eff}$=4500 K, they
determined a lithium abundance of log~N(Li)=0.4 where 
log~N(Li)=12+log~(N(Li)/N(H)).

We carried out a fine spectroscopic analysis to determine the effective temperature 
of Gl879.  The equivalent widths of 55 clean unblended FeI lines were measured 
from our echelle spectrum.   Solar gf-values were previously determined for
these lines using solar equivalent widths measured from the Kurucz et al. (1984) 
solar atlas and the Kurucz solar atmospheric model (Kurucz 1992).  The details and
full list of gf-values will be provided in Balachandran, Carr and Lambert (1996).
By requiring that the Fe abundance be constant for lines having different   excitation
potential, the spectroscopic temperature estimate is 
4620$\pm$150 K.  The Fe abundance of Gl879 is [Fe/H]=--0.11$\pm$0.02.
A spectral synthesis of the 6707.8 \AA{ } region was performed with the LiI, 
FeI and CN features to obtain a Li abundance of log~N(Li)=0.6$\pm$0.15 (see
Fig. 1b).

In order to compare the Li abundance of Gl879 to that for the open cluster
stars, we must first make certain that the conversions from the observables
to T$_{\rm eff}$ and Log N(Li) used for Gl879 are compatible with those used
 for the
cluster stars.  Since a model atmosphere temperature analysis such as we have
done for Gl879 is not possible for all of the cluster stars, we need to
see how our temperatures and abundances would change if we 
used color-temperature
conversions appropriate for the open cluster stars.  As one test, we have
derived a new T$_{\rm eff}$ estimate for Gl879 using the ad hoc "tuned" 
color-temperature conversion advocated by Stauffer et al. (1995) based on
a comparison of the Pleiades single star V vs. (V-I)$_{\rm C}$ main 
sequence to the D'Antona
and Mazzitelli tracks.  With this temperature --Teff=4440 K--
and Soderblom et al's (1993b) curves of growth, we obtain log N(Li) = 0.5.
In order to check the uncertainties introduced by the color-temperature
conversion, as a second test we have estimated Gl879's  effective
temperature using the (B-V)--T$_{\rm eff}$ calibration adopted by Thorburn et
al. (1993).  This is exactly the temperature scale we use in Section 3
for the open cluster stars.  With this scale, we derive
 T$_{\rm eff}$=4500 K, which
then yields log N(Li) = 0.6 with Soderblom et al.'s (1993b)
curves of growth. 
These results do not differ significantly from our rigorous temperature
determinations, so we adopt T$_{\rm eff}$=4500 K, log N(Li)=0.6 as a suitable
compromise.



\section{Discussion}

In the following discussion, we will attempt to age-date Gl879,
and therefore Fomalhaut's age,
by comparison of its properties to those of stars in several open
clusters.  For this to provide an age for Gl879, we must adopt
ages for the open clusters; there is uncertainty in this step,
however, as there is often disagreement over the exact age to
attribute to a given cluster.  We have attempted to adopt a
consistent age scale for the set of clusters to which we will
compare our Gl879 data after having made a survey of the relevant
literature (Patenaude 1978; Giannuzzi 1979; Mermilliod 1981; 
Meynet et al. 1993; Soderblom \& Mayor 1993; Jones and Prosser
1996).  For definiteness, the age scale we adopt is:  Pleiades -
85 Myr; M34 - 200 Myr; Ursa Major - 300 Myr; Hyades - 700 Myr.
Our derived age for Gl879 is directly tied to this age scale; if
future efforts provide better ages for these clusters, then our
age estimate for Gl879 should be adjusted appropriately.

\subsection{The Age of Gl879  as Estimated from Its Lithium
Abundance}


Figure 2a shows the
 lithium abundance against the effective temperature for the 
Hyades (filled circles for single stars and wide binaries and filled
triangles for tidally-locked binaries) 
and Pleiades (open circles) open clusters.
The original lithium equivalent widths for the Hyades
 were selected from  Boesgaard \& Tripicco
(1986),  Rebolo \& Beckman (1988),
Soderblom et al. (1990), Thorburn et 
al. (1993) and  Barrado y Navascu\'es \& Stauffer (1996).
The abundances of Pleiades stars were taken from 
Pilachowski et al. (1987),
 Soderblom et al. (1993b) and  Garc\'{\i}a--L\'opez et al. (1994).
We estimated the lithium abundances using the Thorburn et al. (1993)
 temperature scale
and  curves of growth used  for Gl879.  
There is some evidence that tidally-locked binaries may retain a
larger fraction of their initial lithium abundance perhaps due to the absence of
rotational spin-down (Zahn 1994).  Comparing Gl879 with the {\it single} Hyades
stars, it is immediately clear that Gl879 must be considerably younger
than the Hyades.  Based on this figure alone,
Gl879 could be as young as the Pleiades, since it falls along the lower
envelope of lithium abundances for Pleiades stars of the same color.

Although the data are sparse, clusters between the ages of the Pleiades
and the Hyades, the UMa Moving group at 300  My
  and M34 at 200  Myr, provide further constraint.  
Figure 2b shows the available lithium data in the UMa Moving Group (filled circles, 
Soderblom et al. 1993a) and M34 cluster (open circles, Soderblom 1995).  
Gl879 appears to fall along the lower envelope of M34.  The overlap
between lithium abundances in M34 and the Pleiades and the scatter in abundances
at a given T$_{\rm eff}$ preclude any firm estimate of the lower age limit for Gl879
from the lithium data.  (A stronger constraint on the lower age limit for Gl879 will 
be obtained from X-ray fluxes in Section 3.2.)  Comparison of the UMa and
M34 abundances shows that UMa abundances are consistently lower than M34
both around 6000 K and 5000 K.  A lithium depletion curve (age = 300 Myr, initial
rotational velocity equal to 10 km~s$^{-1}$) from Chaboyer (1993) is plotted to indicate
the general shape of lithium depletion as a function of T$_{\rm eff}$.  Although there is 
only a single upper limit
in the UMa data at the temperature of Gl879, we extrapolate the available
data to arrive at the reasonable speculation that UMa abundances will probably 
lie below Gl879 at 4500 K.  Given this hypothesis, we conclude that 
the age of Gl879 is less than 300 Myr.

\subsection{The Age from the X-ray Activity}



The rotational velocities of low mass stars decrease with time due to
angular momentum loss from stellar winds.  However, particularly for
stars older than about 100 Myr, these rotational velocities become
difficult to measure because they drop below the resolution limit
of standard high-resolution spectrographs.  The chromospheric
and coronal activity of low mass stars also decrease with increasing
age, presumably as a direct result of the decreasing rotational
velocities, and thus it is possible to also use measures of these
properties as age indicators.  
 In some cases, these surrogate
  rotational velocity indicators provide better constraints than the
  rotational velocities themselves because it is easier to measure the
  surrogate indicators than to measure rotation directly.

Figure 3a shows Log Lx versus (B--V) for Pleiades members (Stauffer
et al. 1994) and for Gl879 (Panagi and Mathiotakis 1993), and Figure 
3b shows similar data for the Hyades (Stern et al. 1995).  Filled
circles represent actual values, and open triangles represent upper
limits; the cross indicates the value for Gl879.  The most important
conclusion to be drawn from these plots is that Gl879 must be
{\it significantly older} than the Pleiades since its coronal activity -
and thus presumably its rotational velocity - is much less than that
for any Pleiades stars of the same color.  The X-ray data for the Hyades
do not provide a very useful constraint - Gl879 could be as old or even
slightly older than the Hyades, because there is a large
spread in the X-ray emission of Hyades stars at a given mass.
We thus conclude that the X-ray data indicates that the age
of Gl879 is significantly larger than 85 Myr.


\subsection{The Age as Estimated from Rotation}

As mentioned in the previous subsection,
the rotation of solar-type main sequence stars decreases with increasing
age, so that the rotational period can also be used as an age indicator.
Actual rotational periods versus the (B--V) color indices are 
plotted in Fig. 4. The data were
selected
from Prosser et al. (1993, 1995) --Pleiades, open circles-- and Radick
et al. (1987) 
--Hyades, filled circles. 
Rotational periods are available only for a small fraction of the
stars in each cluster.  For the Hyades, this is not a problem because
in the relevant color range all Hyades members are likely to follow
a very tight period-mass relation (Duncan et al. 1984); for the Pleiades
however, the selection effects are important since it is much easier to
derive rotational periods for rapid rotators than for slow rotators.
Based on the vsini data available in the literature
(see Stauffer et al. 1994 and references therein),
 only one quarter to one third of the Pleiades
K dwarfs should have periods less than one day 
whereas the majority of them have
vsini $<$ 10 km~s$^{-1}$ and thus should have periods of 5 days or longer.
Therefore, the lack of
Pleiades members with P $>$ 4 days and (B--V) $>$ 1.1 
in Figure 4 is entirely
due to selection effects in the photometric monitoring programs.  Thus,
we interpret Figure 4 as indicating that Gl879 is very probably 
younger than the Hyades and older than the Pleiades, though the sparseness
of the rotational velocity database (particularly the lack of period 
determinations for the slowly rotating Pleiades K dwarfs) 
limits our confidence in these
limits.  Efforts currently in progress to determine rotation periods of
Pleiades and Hyades K dwarfs (Krishnamurthi et al. 1995; Allain and
Bouvier 1995) should allow an improved estimate in the near future.


\subsection{The Age as Derived from the Color--Magnitude 
Diagram}

We have tried to estimate the age of Gl879 using its position on
the
 M$_{\rm V}$--(V--I)$_{\rm C}$ plane, following
 Stauffer et al. (1995). Figure 5 shows the D'Antona \&
Mazzitelli
(1994) isochrones for 3, 10, 35, 70 Myr and the Zero
Age Main Sequence (ZAMS).  
The Pleiades data are shown as open circles.
Based on this diagram alone, Gl879 would be assigned an age slightly
older than 35 Myr.  However, for a variety of reasons, we do not believe
that Gl879 can be this young.  Plausible errors in the observational
properties can shift Gl879 to much older ages in this diagram.  For
example, if Gl879 was really at the distance of Gl881 (allowed within
the two sigma errors), then Gl879 would be 0.4 magnitudes fainter in
Figure 5 and would be essentially on the ZAMS and thus compatible with
any age up to several Gyr.  The position of Gl879 could also be shifted
in the diagram due to spot-related photometric variability. 
For these reasons (errors in the observational properties, spot--related
photometric variability), we believe that, in this particular case, this
diagram is not accurate enough for our purposes. 


\subsection{ Summary of Age Constraints for Gl879}

Table 2 summarizes our attempts to constrain the age of Gl879 (and hence
to constrain the age of Fomalhaut).  Lithium provides the best constraint
on the maximum age of the system, indicating that Gl879 is less than about
300 Myr old.  Coronal activity provides the best constraint on how young
the two stars could be - indicating that they are significantly 
older than 85 Myr.
Thus, our age estimate for Gl879 and Fomalhaut is 200$\pm$100 Myr.

\section{Conclusions}

Based on different properties of 
the late spectral type companion of Fomalhaut,
 we have estimated the age of this Beta Pic star,
yielding a value of 200$\pm$100 Myr. This age is totally
compatible with, and independent of, the age derived for the
primary from its position in the CM diagram and thus adds confidence to
post-ZAMS CM diagram ages derived for the other Beta Pic prototypes
by Backman and Paresce.  The comparison between the ages and IR
excesses between Fomalhaut and HR4796A suggests that the disks evolve
with time, as expected.

As an aside, we note that Anosova \& Orlov (1991) have suggested that
Gl879 and Gl881 may be part of a moving group including the
multiple system ADS6175, which contains three spectroscopic binaries
(Castor A, Castor B and YY Gem).  This moving group has been proposed to
contain about 18 stars, having spectral types between A1V and M6Ve.
The positions of several members of the group in Mv-(V-R)$_{\rm C}$ 
and Mv-(B--V)
diagrams agree with our age assignment for Gl879.

If this moving group is real and Fomalhaut and Gl879 are part
of it, all these stars would have the age estimated for Gl879.  YY Gem,
a very well known eclipsing binary, is one of these stars.  It is one
of the two systems of M dwarfs with accurate measurements of radii and
temperatures.  Depending on the exact age which is assigned, YY Gem may
be a pre-main sequence star close to the ZAMS.  If it is indeed still
contracting to the ZAMS, then the calibrations of theoretical models
under the assumption that it is a MS stars would be slightly in error.
This points out the need to continue to search for other appropriate dM 
binaries from which masses and radii can be derived.

Recently, Chabrier \& Baraffe (1995) have estimated the age of YY Gem
by fitting isochrones which were computed using a new equation of state
and new opacities.  They obtained an age of 100 Myr and a solar-like
metallicity, which implies that the components of YY Gem  should still
be slightly above the ZAMS.  Since the age uncertainty for YY Gem
due to errors in its parallax and effective temperature should not
be larger than about 50 Myr, this would suggest that Gl879 has a younger
age than we have derived if YY Gem and Gl879 are indeed coeval.  Since
this is a more indirect way to estimate the age of Gl879 than the methods
we have used, and since it is possible that the Castor group is not
physically associated with Gl879 and Fomalhaut, we have chosen not
to modify our current age estimate for Fomalhout.  However, this 
subject should be readdressed if better parallaxes, proper motions
 and metallicities for Gl879 and YY Gem can be obtained to assess 
 better the reality of their group membership.



\acknowledgements

This research has made use of the Simbad Data base,
operated at CDS, Strasbourg. DBN thanks the Real Colegio
Complutense
at Harvard University for its fellowship. JRS acknowledges
support
from NASA Grant NAGW-2698.

\newpage

\begin{table}
\caption[ ]{Data for the physical pair Gl879 + Gl881 (Fomalhault)
}
\begin{tabular}{cccc}
\tableline
                &     Gl879   &   Gl881  & Reference\\
\tableline
Sp. Type        &   K5 Ve      &  A3 V        & a   \\
V               &  6.46--6.48  &  1.16        & b   \\
(B-V)           &  1.10        &  0.08--0.09  & c   \\
(V-R)$_{\rm C}$ &  0.660       &  0.055       & c   \\
(R-I)$_{\rm C}$ &  0.545       &  0.025       & c   \\
(V-I)$_{\rm C}$ &  1.205       &  0.080       & c   \\
Mv              &  7.02        &  2.09        &     \\
Parallax       &0.128$\pm$0.014&0.154$\pm$0.008& a  \\
U (kms$^{-1}$)  &  -4          &  -5          & d   \\
V (kms$^{-1}$)  &  -8          &  -7          & d   \\
W (kms$^{-1}$)  &  -4          &  -10         & d   \\
$\mu$ (``/yr)   &  0.360       &  0.372       & e   \\
$\theta$(degree)&  114.6       &  115.6       & e   \\
$\gamma$ (kms$^{-1}$)&9.0      &  6.1-6.5     & e   \\
P$_{\rm rot}$ (d)&10.30        &  $\sim$1     & f   \\
Log Lx (ergs$^{-1}$)& 28.3     &    --        & c   \\ 
\tableline
\end{tabular}
$\,$\\
a SIMBAD database.
b Oudmaijer et al. 1992.  
c Panagi $\&$ Mathioudakis 1993.
d Anosova $\&$ Orlov 1991.
e Poveda et al. 1994.
f Busko \& Torres 1978.


\end{table}

\newpage

\begin{table}   
\caption[ ]{Different estimations for the age of Gl879.}
\begin{tabular}{ccc}
\tableline
Method    &     Age     & Quality \\
          &   (Myr)     &         \\
\tableline
Lithium   &   $<$300    & Good    \\  
Activity  & $>$85,$<$600& Good    \\
Rotation  & $>$85,$<$600& Fair    \\
CM diagram&    $>$30    & Poor    \\
          &             &         \\
Final     & 200$\pm$100 &  --     \\       
\tableline
\end{tabular}
\end{table}

\clearpage

\begin{center}
{\sc\large Figure Captions}
\end{center}

\figcaption{{\bf a} Spectrum of Gl879. As can be seen, FeI6707.4
\AA{ } and LiI6707.8 \AA{ } are resolved, due to the high
resolution  and to the small {\it vsini} value. 
{\bf b} Detail around LiI6707.8 \AA. The original spectrum is shown as 
a solid line whereas the synthetical fit appears as a dotted line. 
\label{fig1}}

\figcaption{Li abundance against effective temperature. 
{\bf a} Hyades (filled circles represent single stars and normal 
binaries and filled triangles represent tidally-locked
binaries) and Pleiades (open circles)
data. 
{\bf b} UMa Group (filled circles) and M34 cluster (open
circles). A lithium depletion isochrone by Chaboyer (1993),
computed with an initial rotational  velocity equal to 10 km~s$^{-1}$ 
and an age of 300 Myr,  is shown. \label{fig2}}

\figcaption{ Luminosity in X rays -in logarithm- against the
color index (B--V). Filled circles represent actual values,
whereas open triangles are upper limits.
{\bf a} Pleiades data. {\bf b} Hyades data. \label{fig3}}

\figcaption{Comparison between the rotational periods of
Gl879, Hyades stars (filled circles) and Pleiades members (open
circles). \label{fig4}}

\figcaption{Color--Magnitude Diagram for Gl879, HR4796B 
-physical companion of another Beta Pic analog- and the Pleiades
(open circles). The isochrones by D'Antona and Mazzitelli
(1994), for ages 3, 10, 35, 70 Myr and the ZAMS, from top to bottom,
 are also included. \label{fig5}}


\begin{figure}
\vspace{18cm}

\includegraphics{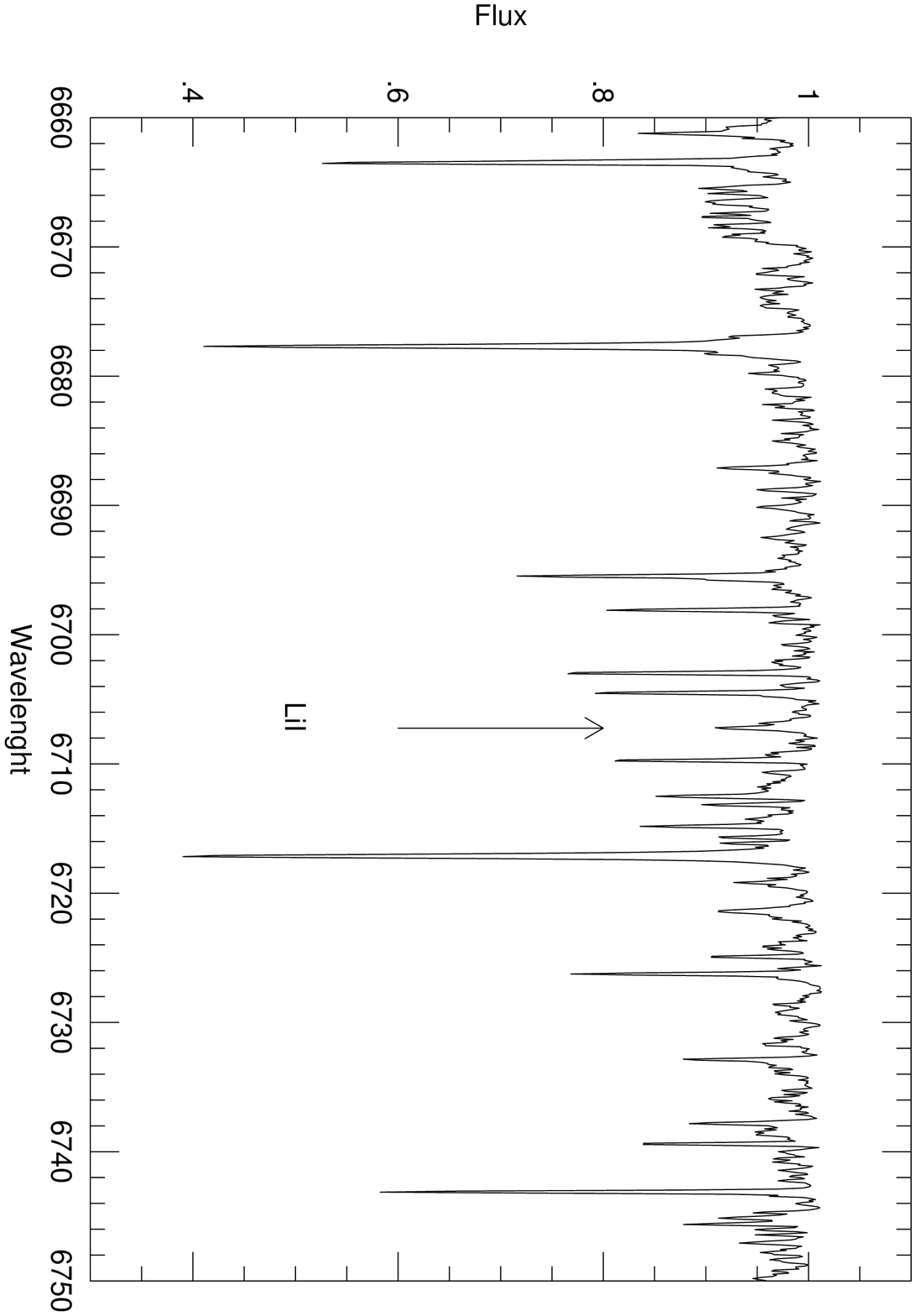}

\end{figure}

\begin{figure}
\vspace{18cm}

\includegraphics{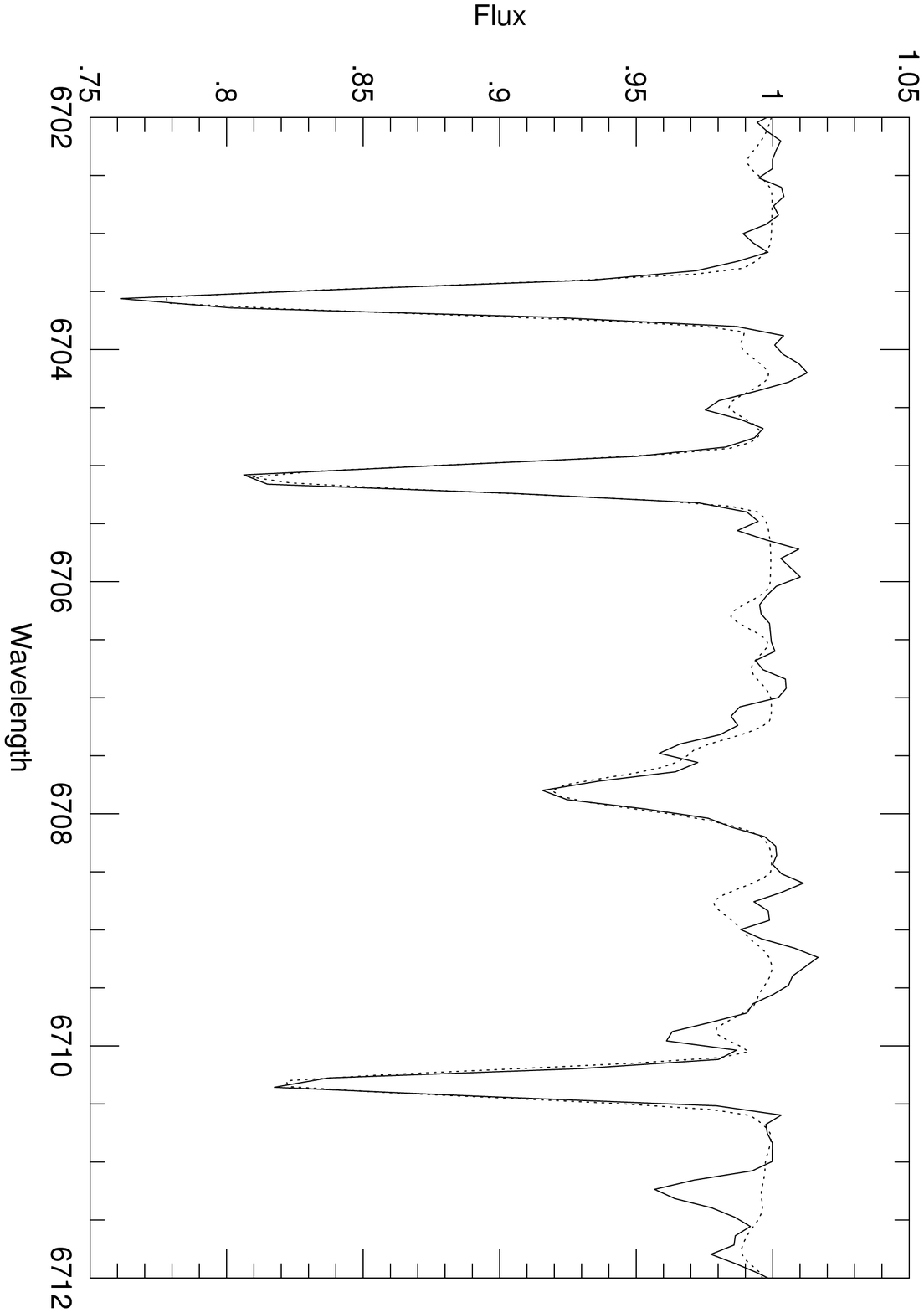}

\end{figure}

\newpage

\begin{figure}
\vspace{18cm}

\includegraphics{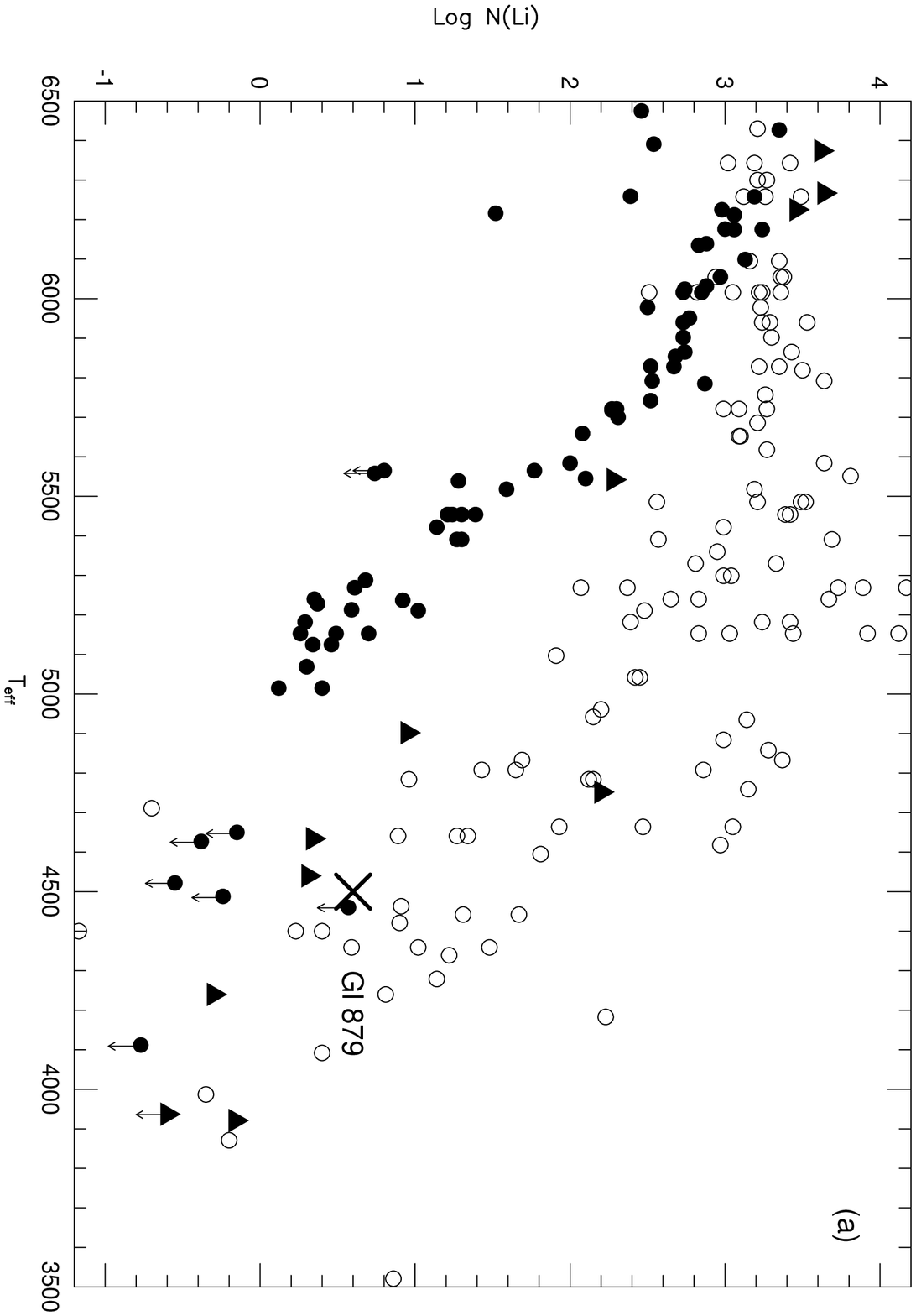}

\end{figure}
\newpage

\begin{figure}
\vspace{18cm}

\includegraphics{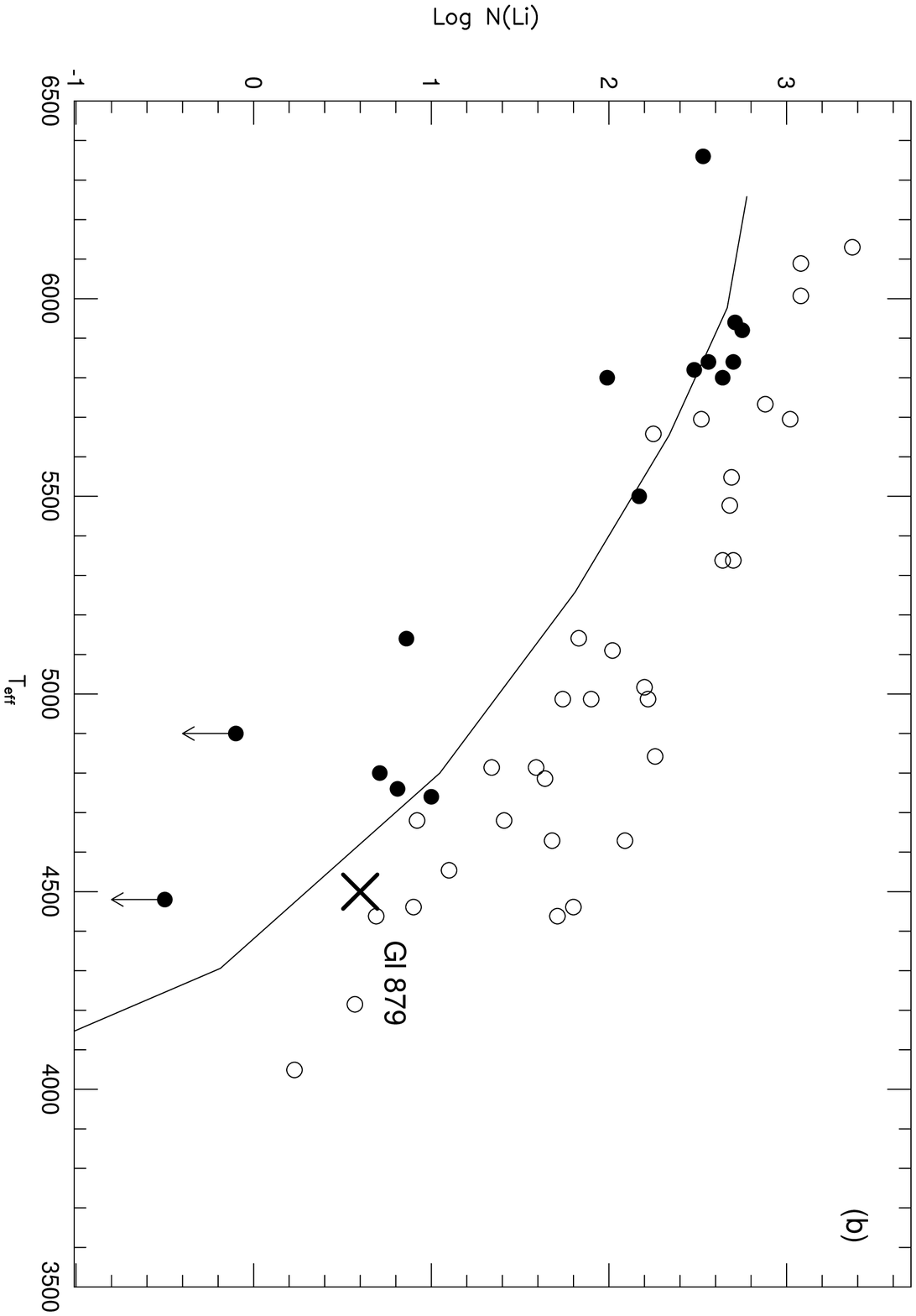}

\end{figure}
\newpage

\begin{figure}
\vspace{18cm}

\includegraphics{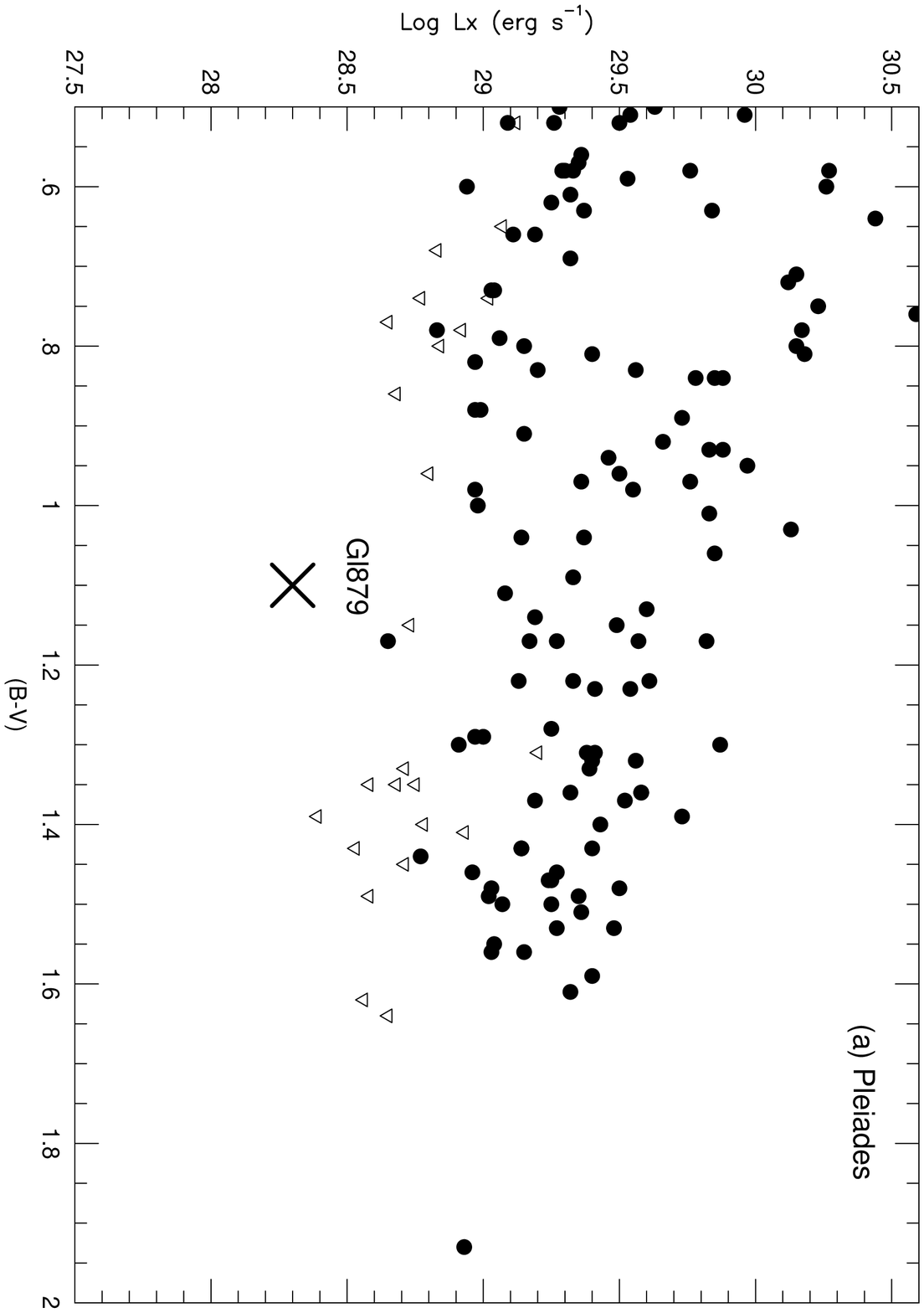}

\end{figure}
\newpage

\begin{figure}
\vspace{18cm}

\includegraphics{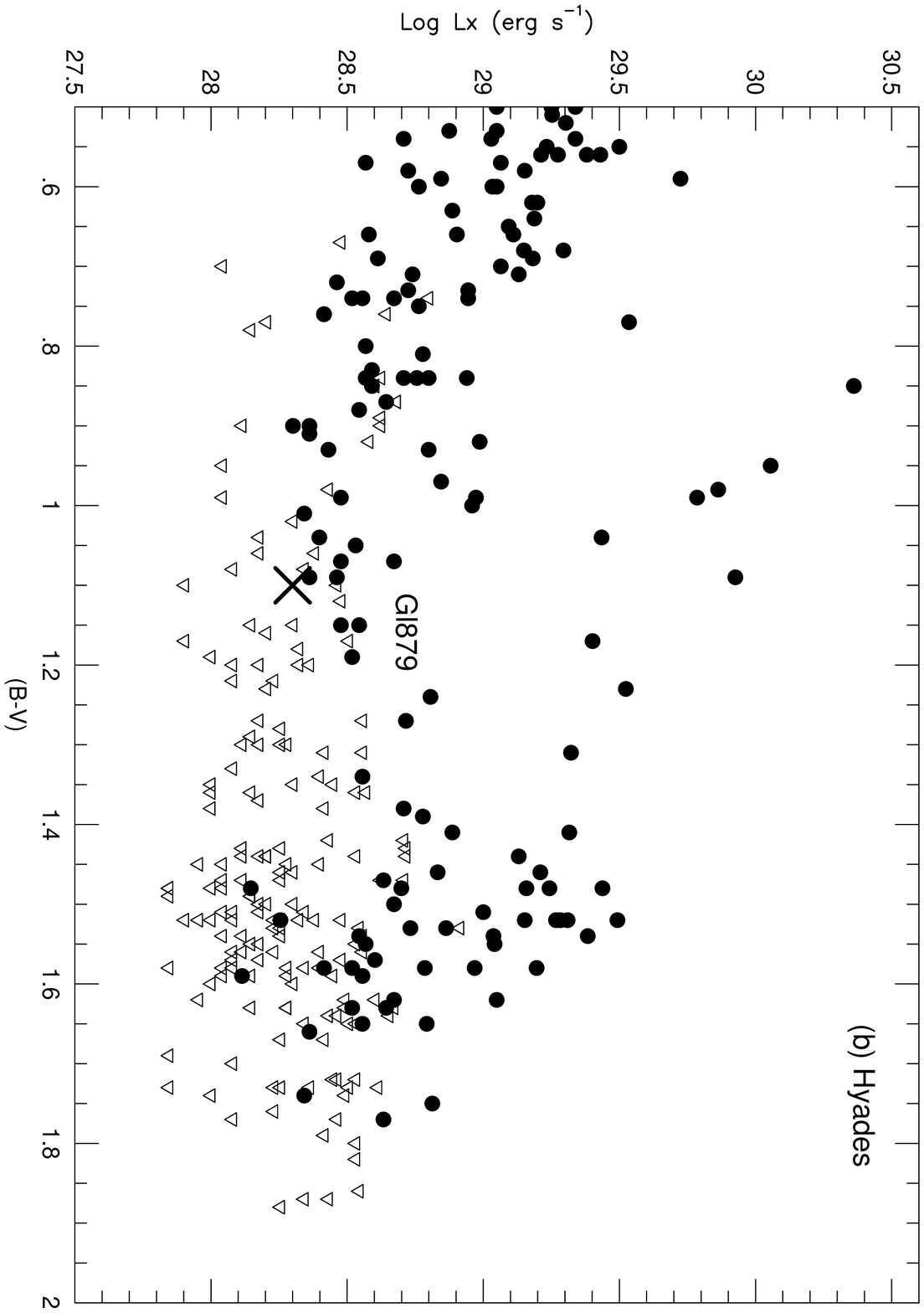}

\end{figure}
\newpage

\begin{figure}
\vspace{18cm}

\includegraphics{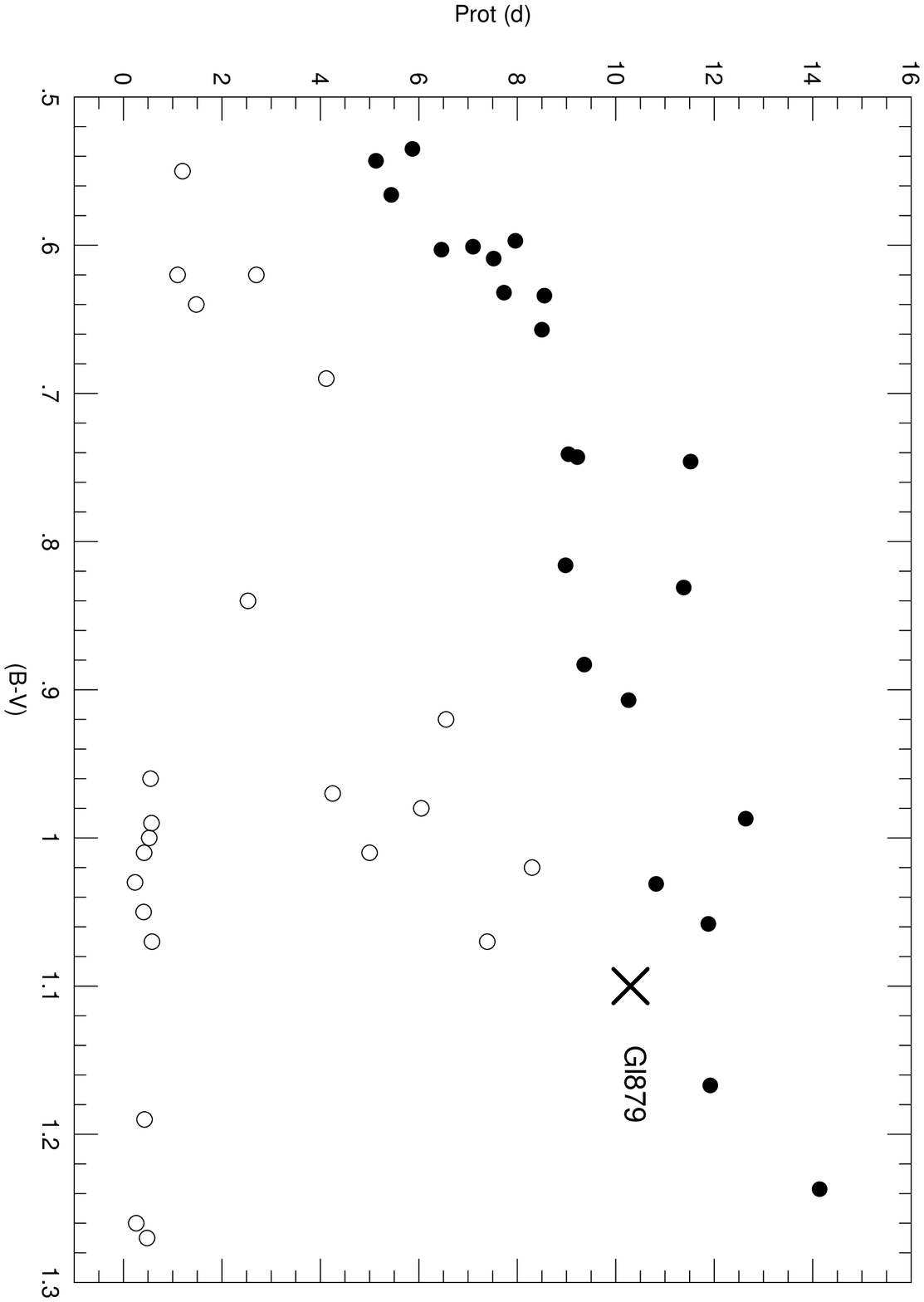}

\end{figure}
\newpage

\begin{figure}
\vspace{18cm}

\includegraphics{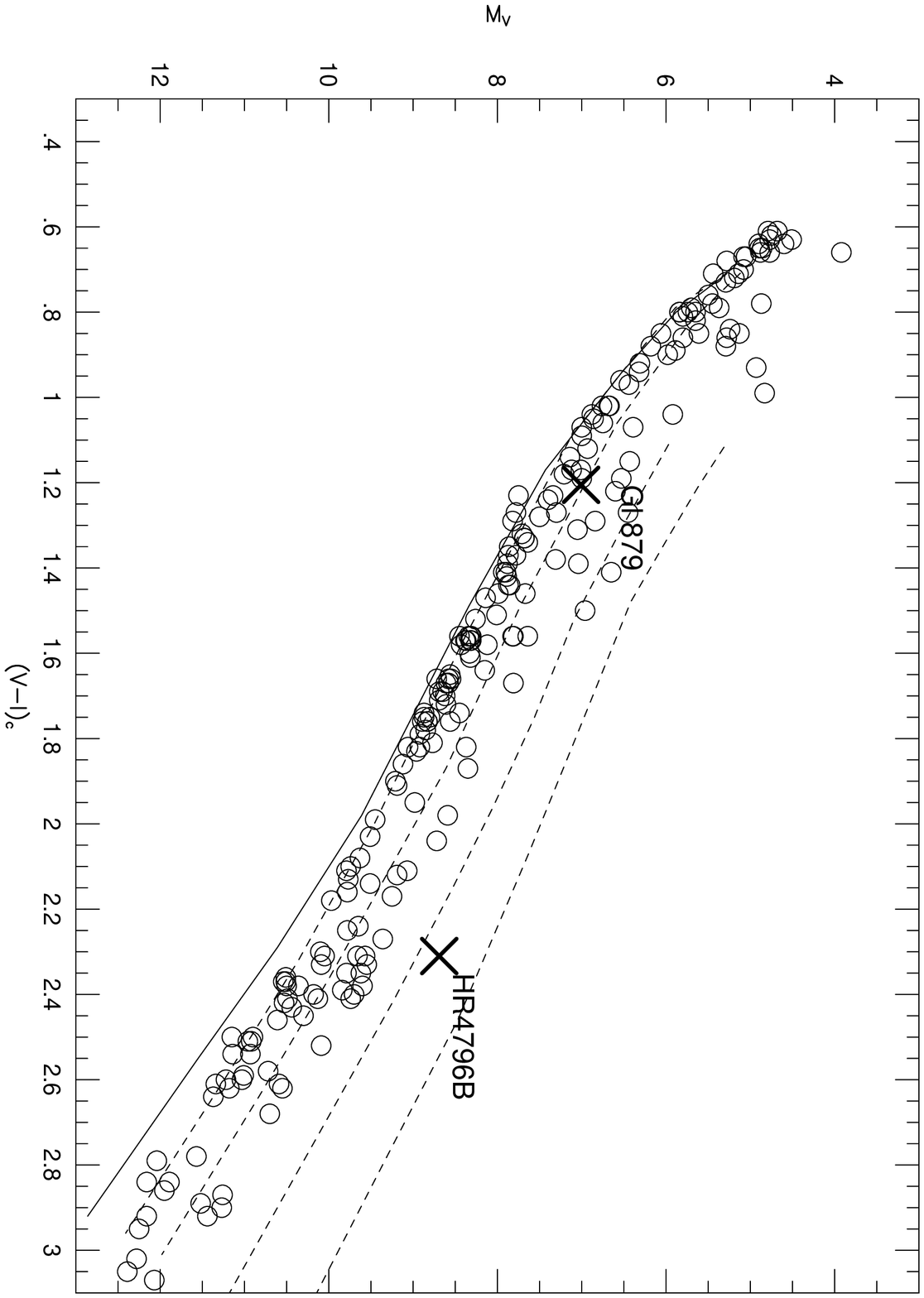}

\end{figure}

\end{document}